\documentclass[journal=jpccck,manuscript=article]{achemso}

\usepackage{hyperref}
\usepackage[latin1]{inputenc}
\usepackage{graphicx}
\usepackage{dcolumn}
\usepackage{bm}
\usepackage{ae}
\usepackage{aecompl}
\usepackage{amsmath}
\usepackage{amssymb}
\usepackage{epsfig}
\usepackage{psfrag}
\usepackage{color}
\usepackage[final]{pdfpages}

\renewcommand{\ell}{h}

\author{Dariusz Tarasewicz}
\affiliation{ Department of Theoretical Chemistry, Institute of Chemical Sciences, Faculty of Chemistry, Maria Curie-Sklodowska University in Lublin, Lublin, Poland.}
\author{Edyta Raczy{\l\l}o}
\affiliation{ Department of Theoretical Chemistry, Institute of Chemical Sciences, Faculty of Chemistry, Maria Curie-Sklodowska University in Lublin, Lublin, Poland.}
\author{Wojciech R{\.z}ysko}
\affiliation{ Department of Theoretical Chemistry, Institute of Chemical Sciences, Faculty of Chemistry, Maria Curie-Sklodowska University in Lublin, Lublin, Poland.}
\author{{\L}ukasz Baran}
\affiliation{ Department of Theoretical Chemistry, Institute of Chemical Sciences, Faculty of Chemistry, Maria Curie-Sklodowska University in Lublin, Lublin, Poland.}
\email{lukasz.baran@mail.umcs.pl}

\title{Self-assembly of chromatic patchy particles with tetrahedrally arranged patches}

\begin{document}

\begin{abstract}The achievement of selectivity in the formation of cubic diamond is challenging due to the
 emergence of competing phases such as its hexagonal polymorph or clathrates possessing similar free energy. 
 {\color{black} Although both polymorphs exhibit a complete photonic bandgap, the cubic diamond exhibits it at lower frequencies than the hexagonal counterpart, positioning him as a promising candidate for photonic applications.}
 {\color{black} Herein, we demonstrate that
 the 1:1 mixture of identical patchy particles cannot selectively form the cubic diamond polymorph due to the frustrations present in the system that are manifested in the primary adsorption layer and propagate as the film grows}. We provide {\color{black} a plausible} explanation why the binary system under confinement, resembling interactions between the complementary DNA bases cannot yield the selectivity in the formation of cubic diamond crystals {\color{black} which is based on the similarities to the antiferromagnetic systems}. 
 We always observe
 the mixture of both hexagonal and cubic diamonds, however, the formation of such stacking hybrids is observed for a wider range of patch sizes compared to the one-component system. 
\end{abstract}

\section{Introduction}

Patchy particles are a promising class of colloidal building blocks. They often find an application for the design of various ordered networks at the microscale \cite{Duguet2017, Duguet2020}. A characteristic feature of patchy particles 
is the presence of interaction areas that can be achieved by specific patterning of its surface. 
Therefore, contrary to the isotropic hard spheres that form only close-packed networks due to excluded volume effects \cite{Poon2004}, they are ideal candidates for obtaining open lattices, so-ought due to their predicted applications in photonics \cite{photonic_appl, Cersonsky21}, as sensors, semiconductors, and many others \cite{Travesset2024}. It is because controlling the shape, directionality of interactions, and other properties, all of which have been extensively studied using computer simulations for many years \cite{Tetraedr_PNAS, Noya2019, Nano_rods2021, Noya1, Romano2012, sanz, sanz1, zhang1, dijkstra}, is now achievable also experimentally \cite{Ducrot2017, He2020, Science2016, Pine2021, Liedl2024, Rovigatti2024}. This allows for the formation of diamond, pyrochlore (tetrastack), or clathrate networks, depending on the delicate manipulation of the aforementioned parameters.

One would expect that the most straightforward approach relies on arranging the patches in such a way that they should enforce how the particles in a target lattice are distributed. The most prominent example of why such a scheme alone is not sufficient can be depicted by tetrahedral patchy particles. In the bulk conditions, they indeed form diamond networks, however, the selectivity in the formation of a desired polymorph cannot be achieved due to the small free energy difference between the hexagonal (HD) and cubic diamond (CD) phases \cite{sanz, sanz1}. In such a case, the formation of crystals with interwoven CD/HD phases is observed, instead. On the other hand, narrowing the size of the patches, effectively increasing the directionality of interactions, results in the emergence of empty sII clathrate cages \cite{Noya2019}, isostructural to ice XVI \cite{Falenty2014}. 
This suggests that not only the patch's arrangement is the key, but the excluded volume effects stemming from different shapes (and the interplay between these two aspects) are also important to obtain target crystalline structures.

In fact, to date, the experimental formation of a cubic diamond in a single-component system has been reported using tetrahedral, non-spherical clusters \cite{He2020}. On the other hand, the formation of clathrate networks from patchy particles is still to be achieved  \cite{Pine2021}, despite theoretical predictions \cite{Romano2012, Noya2019}. However, it has been shown that hard polyhedra of different sizes tend to form such structures \cite{Glotzer_klatrat, Glotzer_klatrat2}, which can be useful for guest-host molecule reactions. 

Another possibility to achieve the target network is to use templates strictly commensurate with the growing structure on top of it.  
In this sense, Davies \textit{et al.} demonstrated the possibility of the formation of a metastable cubic ice polymorph via heterogenous nucleation \cite{IceIc_Michaelides}. 
Subjecting the particles to the external field can also drive the formation of a given crystalline network. In fact, the CD polymorph was found to be formed when the electric field was applied to water molecules \cite{Niall2018}. 
These observations are expected as it is widely accepted that such an additional degree of freedom can significantly alter the phase diagram \cite{Koga2001, Koga2002}. 
Recent advances in this field were described in an excellent review article \cite{dijkstra_perspective}.

In our latest works \cite{Baran2023, Baran2024}, we have demonstrated that tetrahedral patchy particles can assemble into a cubic diamond selectively when the system is subjected to a sufficiently strong external field modelled using a Lennard-Jones (9,3) potential. The driving force of such a nucleation mechanism relied on the formation of a first adsorption layer being strictly commensurate with the (110) face of the cubic diamond polymorph. Since the stacking hybrids are formed only when the (111) face of the CD polymorph is exposed, the selectivity has been achieved. Although these results are significant due to the formation of CD polymorph from regular tetrahedral patchy particles in a one-component system, DNA-mediated interactions have been widely used to steer the interactions between colloidal particles \cite{Liedl2024, Rovigatti2024, Wang2012, Rogers2016}.
As the DNA interacts only between complementary bases, a two-component mixture 
needs to be considered. A 1:1 mixture of such identical tetrahedral patchy particles with mixed interactions has been already studied by Neophytou \textit{et al.} in the bulk conditions 
\cite{Tetraedr_PNAS}. Such a simulation setup is often termed a "chromatic" patchy particle system and is expected to increase the stability of crystalline phases \cite{chromatic2015}.

{\color{black}Notwithstanding recent advances, we continue the investigations of tetrahedral patchy particles subjected to external field potential. In this paper, we aim to assess whether the
effects observed in our previous articles \cite{Baran2023, Baran2024} are universal features and could also be observed in a 1:1 mixture of identical patchy particles interacting only between distinct species. We have found that
{\color{black} even though for the strong external field, the majority of the system crystallizes into the CD polymorph, the selectivity in its formation cannot be achieved due to the frustrations present in the system that are manifested 
in the primary adsorption layer and propagate as the film grows.}
Moreover, similar to previous findings \cite{Tetraedr_PNAS}, we demonstrate that (i) metastable empty s-II clathrate networks are never formed both in the bulk conditions and upon subjecting particles to external field 
and (ii) diamond networks emerge for a wider range of patch's width, in comparison to one-component systems.}

\section{Methods}
\subsection{Patchy particle model}

\begin{figure}[h!]
    \centering
    \includegraphics[width=0.9\linewidth]{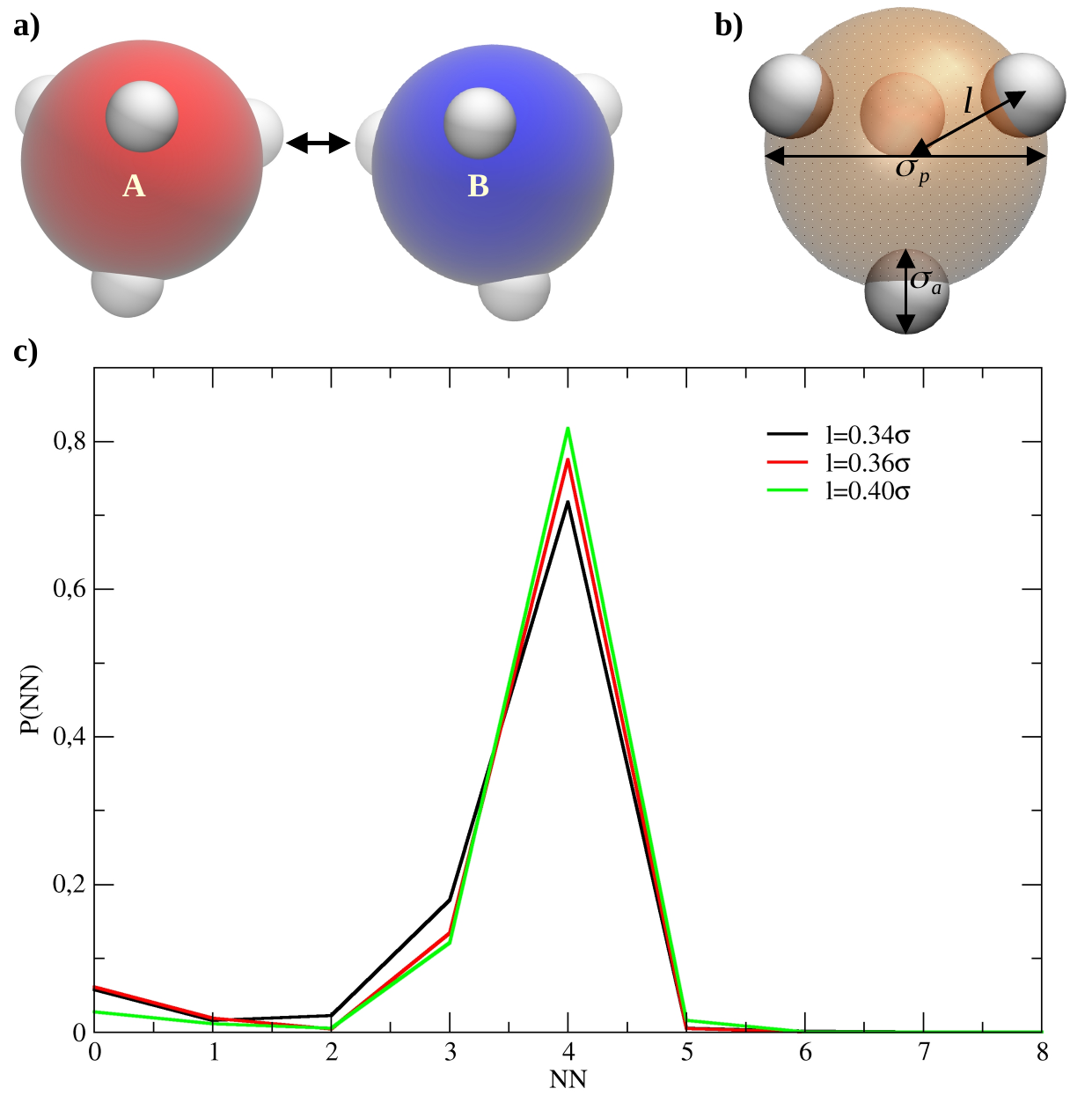}
    \caption{Part (a): Model of tetrahedral patchy particles of two different kinds A (in red) and B (in blue). The active sites are marked in white.
    Part (b): The description of the model's parameter. The scale is not preserved. 
    Part (c): The distribution of nearest neighbours for different
    embedding distance $l$. }
    \label{fig:model}
\end{figure}

We consider a 1:1 mixture of identical tetrahedral patchy particles of kinds A and B as shown in Figure~\ref{fig:model}-a. The particles interact only between distinct species. As already mentioned, this can be achieved experimentally using DNA-mediated interactions \cite{Liedl2024, Rovigatti2024, Wang2012, Rogers2016}. 
They are comprised of a central spherical core of size $\sigma_p$ to which surface four active sites, each of size $\sigma_a$, were embedded to a certain extent (Figure~\ref{fig:model}-b). The latter can be manipulated by a parameter $l$. In the present study, we have considered three cases: (i) narrow $l=0.34\sigma$,
(ii) intermediate $l=0.36\sigma$, and (iii) wide patches $l=0.40\sigma$. However, it is noteworthy
that these three cases are still within the one-bond-per-patch regime, as can be anticipated from 
Figure~\ref{fig:model}-c, since the most probable number of neighbours for a given
particle is equal to four. 
The patchy particle's geometry is maintained using the harmonic potential for both bonds and bond angles. 

The patchy particles interact via the truncated and shifted Lennard-Jones (12,6) potential, in which there is no discontinuity in both the potential and the forces \cite{toxvaerd}.

\begin{align}
    U(r)&=u_{LJ}(r)-u_{LJ}(r_c)-(r-r_c)\frac{du_{LJ}}{dr} \Big\rvert_{r=r_c} \\  \nonumber
    u_{LJ}(r)&=4\varepsilon \left [ \left(\frac{\sigma}{r} \right )^{12} 
    - \left(\frac{\sigma}{r} \right )^{6} \right ]
    ~~~~~~~{\rm if~} r\leq r_{cut}
\end{align}

\noindent However, the only attraction in the system is set between the active sites of distinct species of patchy particles. The remaining interactions were soft-repulsive (between units of the same type and cores of different kinds), meaning the potential is truncated at a distance equal to the diameter of a given interaction pair.

To introduce the effects of spatial constraints and enforce the directionality of the growth of emerging networks, we introduce
an external field in the form of the Lennard Jones (9,3) 
potential acting solely on the core of the patchy particle of both kinds A and B.

\begin{equation}
 U^{\mathrm{ext}}(z)=\varepsilon_{wc} \left [\frac{2}{15} \left (\frac{\sigma}{z} \right)^9 - \left (\frac{\sigma}{z} \right)^3 \right ] 
 ~~~~~~~{\rm if~} z\leq5\sigma
\end{equation}

\noindent where $\varepsilon_{wc}$ indicates the depth of potential well. 
To assess whether the cohesion or adhesion prevails, the 
discussion will rely on the parameter $\xi=\varepsilon_{wc}/\varepsilon_{aa}$ depicting the ratio of the strengths of particle-wall $\varepsilon_{wc}$ and associative particle-particle $\varepsilon_{aa}$ energies.
The entire set
of parameters used in this work can be found in Table~\ref{tab:parameters} and Table~\ref{tab:parameters2}. Therefore, it can be easily verified that the model parameters used in this work are identical to those used for one-component systems
in our previous works \cite{Baran2023, Baran2024} and the only difference is that in the present study, we consider a 1:1 mixture of two identical patches with mixed interactions.

\begin{table}[h!]
    \centering
    \begin{tabular}{c|c|c}
    \hline
    \hline
        parameter & symbol  & value \\
    \hline
        core diameter & $\sigma_p$ & $1.0\sigma$ \\
        active site diameter & $\sigma_a$ & $0.2\sigma$ \\
        association energy (between A-B) & $\varepsilon_{aa}$  & $5.0\varepsilon$  \\
        soft-repulsive energies & $\varepsilon_{ij}$  & $1.0\varepsilon$  \\
        association cutoff & $r_{cut,aa}$  & $2.0\sigma_{aa}$  \\
        soft-repulsive cutoffs & $r_{cut,ij}$  & $1.0\sigma_{ij}$  \\
        bond harmonic constant & $k_b$  & $1000\varepsilon/\sigma^2$  \\
        bond angle harmonic constant & $k_\theta$  & $1000\varepsilon/\mathrm{rad}^2$  \\
        \hline
        \hline
    \end{tabular}
    \caption{Parameters of the model. In the above $ij=pp, ap$.
    }
    \label{tab:parameters}
\end{table}

\begin{table}[h]
    \centering
        \scalebox{0.9}{
    \begin{tabular}{c|c|c|c|c}
    \hline
    \hline
        parameter & symbol  & \multicolumn{3}{c}{value} \\
    \hline
 &  & narrow & intermediate & wide \\
embedding distance & $l$  & $0.34\sigma$ & $0.36\sigma$ & $0.40\sigma$ \\
external field energy & $\varepsilon_{wc}$ & $0.5\varepsilon-4\varepsilon$ &
$1\varepsilon-7\varepsilon$ & $1\varepsilon-6\varepsilon$ \\
    \hline
    \hline
    \end{tabular}}
    \caption{Difference between the model's parameters between the narrow, intermediate, and wide patchy particles.}
    \label{tab:parameters2}
\end{table}

\subsection{Crystal identification}

Three-dimensional crystalline environments are detected using the CHILL + order parameter \cite{chill+}. It allows for the distinction between diamond polymorphs and clathrate networks from disordered fluid. The entire idea is based on the correlation function $c_l(i,j)$ defined as 

\begin{equation}
 c_l(i,j)=\frac{\displaystyle \sum_{m=-l}^{m=l} q_{lm}(i)q^*_{lm}(j)}{\left( \displaystyle\sum_{m=-l}^{m=l} q_{lm}(i)q^*_{lm}(i)\right)^{1/2}\left(\displaystyle \sum_{m=-l}^{m=l} q_{lm}(j)q^*_{lm}(j)\right)^{1/2}} 
 \label{eq:chill}
\end{equation}

\noindent where

\begin{equation}
 q_{l}(i)=\sqrt{\frac{4\pi}{2l+1} \sum_{m=-l}^{m=l} \left| q_{lm}(i) \right|^2}
 \label{eq:stein}
\end{equation}

\noindent with

\begin{equation}
 q_{lm}(i)=\frac{1}{N_b(i)} \sum_{j \in N_b(i)} 
 Y_{lm}(\theta_{ij}, \phi_{ij}) 
 \label{eq:stein1}
\end{equation}

\noindent where $Y_{lm}$ are spherical harmonics, and for a given sphere $i$ we choose a set of its nearest neighbors, $N_b(i)$.

It has been found that $c_3(i,j)$ is the best for the discrimination of the above-mentioned networks. 
Two particles are connected by a staggered bond when $c_3(i, j) \leq -0.8$ and by an eclipsed bond when $0.25 \geq c_3(i, j) \geq -0.35$. 
After that, all the molecules with exactly four neighbours were further discriminated as cubic diamonds (clathrate) if they were connected with one another by four staggered (eclipsed) bonds, hexagonal diamond which has three staggered bonds and one eclipsed bond, or interfacial diamonds
with at least two staggered bonds.

\subsection{Simulation details}

Molecular dynamics simulations are performed in the canonical $NVT$ ensemble using the LAMMPS simulation package \cite{LAMMPS}. Trajectories are evolved using the Velocity Verlet algorithm with a time step equal to $\tau=0.001$. The temperature is controlled using the three chains Nos{\'e}-Hoover chains algorithm with the damping factor $\tau_{NH}=10\tau$. Acknowledging that insufficiently large system sizes can promote metastable results due to the emerging frustrations in the system \cite{Baran2023, Baran2024}, we opt for systems comprising of $8100$ ($N_{tot}=16200$) patchy particles of each kind subjected to external field. The system size is set to be $47.8\times47.8\times40$ in the $x, y$, and $z$ directions, respectively, corresponding to the surface density $N_{tot}/(L_x\times L_y)\approx7$ and the total system's density equal to $\rho=0.177$. {\color{black} The periodic boundary conditions are applied only in the $x, y$ directions whereas in the $z-$ direction the box is closed by the attractive (on the bottom) and reflective (on the top) walls.} The system in the bulk is a cubic box comprising of $20^3=8000$ patchy particles of each kind ($N_{tot}=16000$) and is simulated at the density $\rho=0.4$.
{\color{black} This value is chosen so that applying the lever rule, the proportion of two coexisting phases will be almost the same.}

The systems are gradually cooled down from disordered states with a decrement in the temperature
equal to $\Delta T=0.01$. After each run, the detection of ordered networks has been performed
using the $c_3(i,j)$ order parameter. Once the nucleation event happens, the temperature step decreases to 
$\Delta T=0.005$. Simulations are launched
for at least $2\times10^8$ time steps for every thermodynamic state for the equilibration, however,  the simulation time for wide patches ($l=0.40\sigma$) took up to $10^9$ time steps.
Further production runs are launched for at least $2\times10^7$ in order to collect the averages.
Five independent replicas are simulated to check for replica-dependent behaviour. {\color{black} For several cases, fifteen replicas are simulated. }

\section{Results and Discussion}
\subsection{Narrow patches}

\begin{figure*}[h!]
    \centering
    \includegraphics[width=\linewidth]{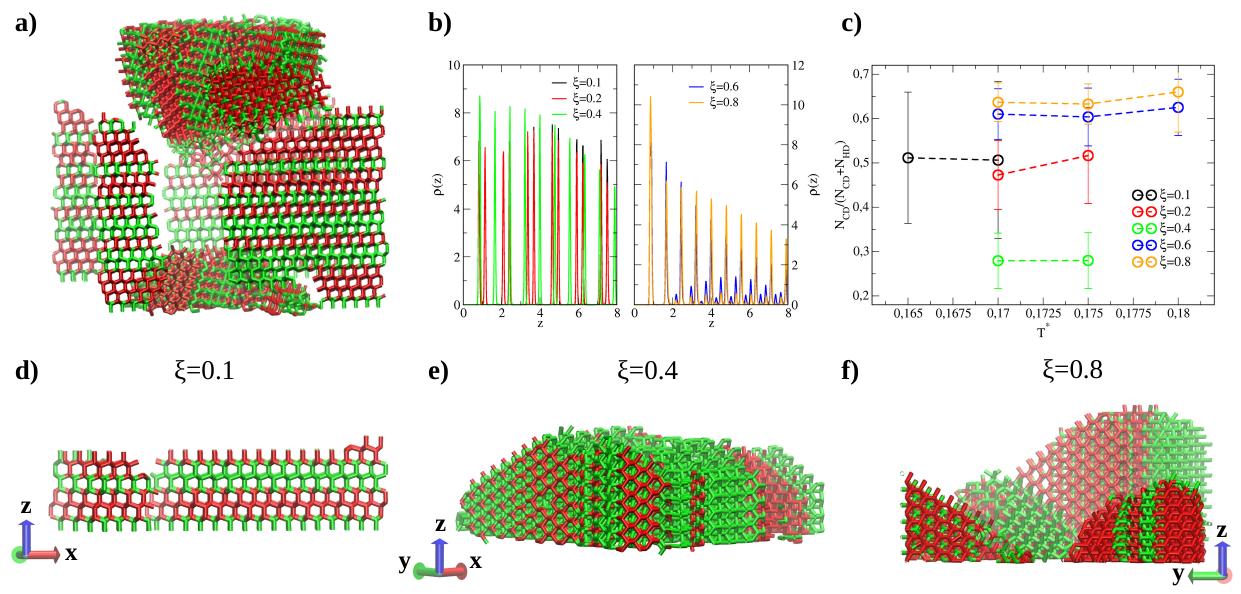}
    \caption{Part (a): Snapshots in the bulk system at $\rho=0.4$ and $T=0.17$. 
    Density profiles (b) and cubicity parameter (c) for the considered systems under different strengths of an applied external field. 
    Parts (d-f): Snapshots demonstrating different crystal growth mechanisms depending on the strength of the surface potential at $T=0.17$.
    Red and green sticks correspond to the cubic and hexagonal diamond environments, respectively.}
    \label{fig:narrow}
\end{figure*}

Let us begin the discussion with the simulation results obtained for narrow patches ($l=0.34\sigma)$ in the bulk. In a one-component system, we have recently observed the emergence of both stacking hybrids of interwoven CD/HD polymorphs and empty s-II clathrate cages \cite{Baran2024}. It is consistent with previous findings by Noya \textit{et al.} \cite{Noya2019}. 
On the other hand, in a two-component system, the emergence of metastable clathrates has never been observed. It can be ascribed to the mixed interactions that result in the formation of only even-membered rings where the first and 
last particles are of different species, as already discussed in \cite{Tetraedr_PNAS}. Since s-II clathrates are comprised of five- and seven-membered rings, their formation has been hindered.
Therefore, the only networks that have been formed are stacking hybrids of CD/HD polymorphs as can be seen in Figure~\ref{fig:narrow}-a. 
{\color{black} It is expected that such interactions resemble antiferromagnetic lattices for which a diamond lattice is not frustrated with nearest-neighbours exchange \cite{antiferromagnet1, antiferromagnet2}.}
The ratio of one diamond form to the other changes from replica to replica.

Almost identical results are obtained when the system is subjected to an external field potential. In the density profiles (cf. Figure~\ref{fig:narrow}-b) for $\xi=0.1$ and $\xi=0.2$ a characteristic bilayer structure can be observed indicating the (0001) or (111) faces of HD and CD, respectively. 
It means that the structures are reminiscent of those observed in the bulk conditions. However, the introduction of the external field facilitates the formation of a crystal. Nevertheless, mixed interactions result in very pronounced grain boundaries. It is visible on the snapshot (cf. Figure~\ref{fig:narrow}-d) that the crystal started either from the CD or HD polymorphs. The composition of the crystalline network that has been formed has been described by the cubicity parameter defined as $N_{CD}/(N_{CD}+N_{HD})$ with $N_{CD}$ and $N_{HD}$ being
the numbers of patchy particles (irrespectively of the type) belonging to the CD and HD polymorphs, respectively. This parameter is plotted in Figure~~\ref{fig:narrow}-c. It is visible that the 
cubicity takes a value around $0.5\pm0.15$, meaning that neither of the polymorphs is favoured over the other. 
{\color{black} The arrangement of each kind of particle
and corresponding density profiles can be found in Figure~S1 and Figure~S2.
The separation of the components in a single layer of a bilayer
can be readily seen, indicating the formation of two single crystals
and grain boundaries (Fig.~S1). 
On the other hand, a single crystal has been formed in another replica (Fig.~S2).
It has to be emphasized that the interactions between different species
are between the layers of a bilayer.}

As the external field is increased to $\xi=0.4$, single peaks start to appear as can be seen in the density profiles (cf. Figure~\ref{fig:narrow}-b). The arrangement of each kind of particle is different in comparison to $\xi=0.2$. There is no separation of the components and the interactions between the particles are within the same adsorption layer. Using the CHILL+ parameter, we have found that this corresponds to the formation of the hexagonal diamond which grows from the $(\bar{1}2\bar{1}0)$ direction as can be seen in Figure~\ref{fig:narrow}-e. Nevertheless, contrary to the one-component system \cite{Baran2024}, the selectivity in its formation has never been achieved (cf. Figure~\ref{fig:narrow}-c). 

Further increase of external field $\xi$ results in that single peaks still appear in the density profile
as the distance from the wall increases, however, the CHILL+ order parameter indicates structural changes. Instead of the formation of HD polymorph or stacking hybrids that are perpendicular to the surface, the growth from the (110) face of CD crystal is observed. 
However, notice that as {\color{black} the distance from the wall,} $z$, increases,
intermediate peaks become visible (cf. Figure~\ref{fig:narrow}-b), which correspond to the stacking hybrids 
forming on the (111) face of the CD polymorph which is exposed by 45 degrees with respect to the surface. 

\begin{figure}[h!]
    \centering
    \includegraphics[width=\linewidth]{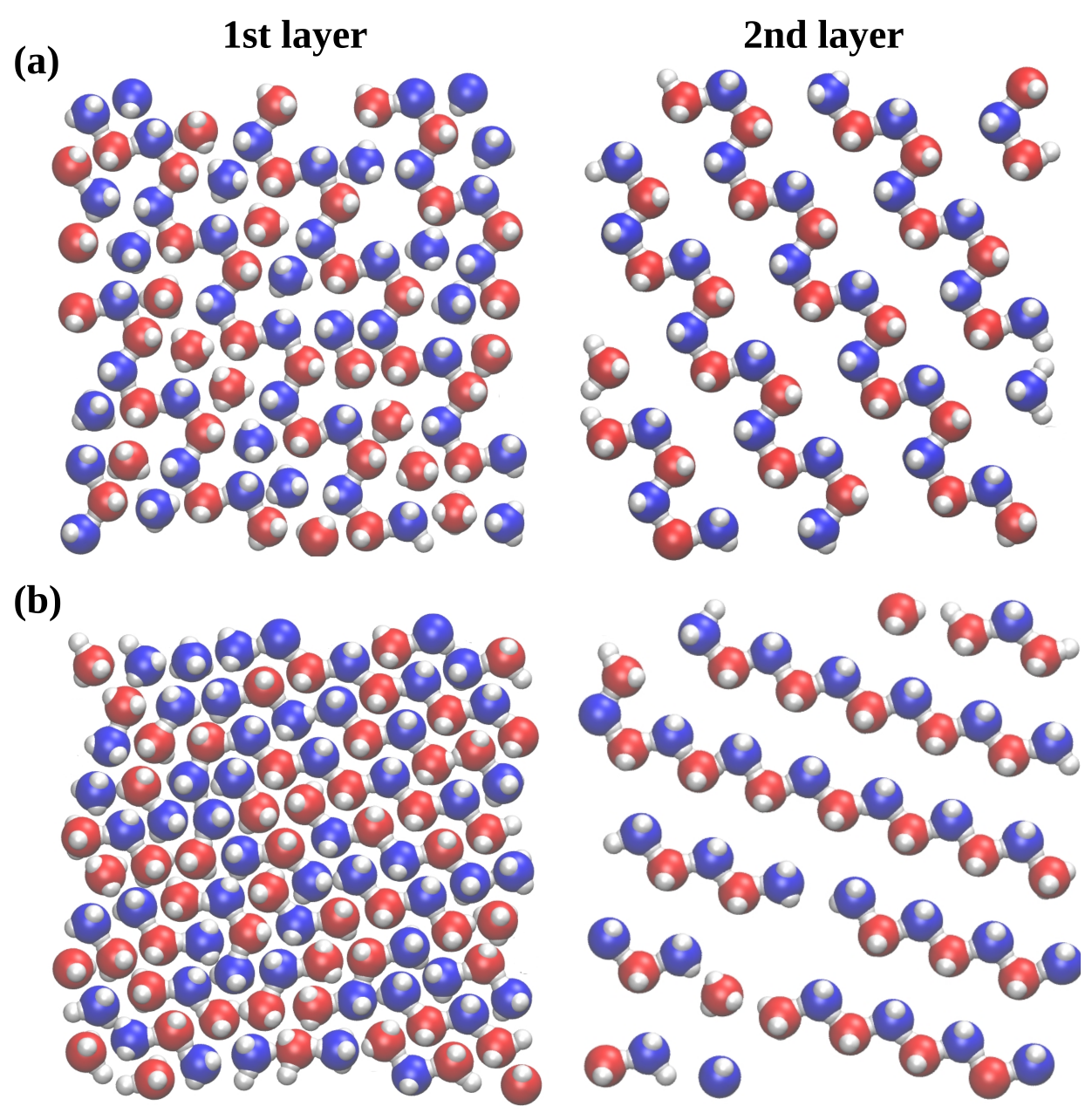}
    \caption{Structures of the first (left-hand side panel) and the second (right-hand side panel) adsorption layers for the hexagonal diamond (a) and cubic diamond (b) formed in the systems with $\xi=0.4$ and $\xi=0.8$ at $T=0.17$, respectively. Patchy particles of different species are marked in different colours. 
    }
    \label{fig:narrow-layers}
\end{figure}

It is demonstrated that such a design that hinders the formation of odd-membered rings results in the formation of only diamond structures and does not allow for the formation of s-II clathrates. 
On the other hand, upon subjecting systems to a sufficiently strong external field potential ($\xi\ge0.4$), does not yield either the HD or CD polymorphs selectively at any conditions (cf. Fig.~\ref{fig:narrow}-c), contrary to the one-component system (cf. Ref.~\cite{Baran2024}). 

WWe conjecture that it is due to the frustrations that are present in the first adsorbed layer that are further propagated to the crystal growing on top of it. As can be seen in the left-hand side Figure~\ref{fig:narrow-layers}-a, in the system with $\xi=0.4$, patchy particles assemble similarly as in the $(\bar{1}2\bar{1}0)$ face of HD crystal (cf. right-hand side panel of Fig.~\ref{fig:narrow-layers}-a), however, mismatches are also visible. 
The same effect can be observed in the system with $\xi=0.8$ as shown in the left-hand side of Figure~\ref{fig:narrow-layers}-b. Despite assembling into a triangular lattice and having the same rectangular unit cell as the (110) face of CD crystal (right-hand side panel Fig.~\ref{fig:narrow-layers}-b), the connections are not so regular as we have found them to be in a one-component system.
{\color{black} Again, an analogy to antiferromagnetic lattices can be readily seen. On the triangular lattice, three neighbouring spins cannot be pairwise antialigned leading to frustration. Here, instead of spins, we have tetrahedral particles that interact only between different species, leading to the same effect \cite{frustration}.}
In consequence, several different monocrystals are formed (Figure~\ref{fig:narrow}-f). Due to such frustrations in the system, CD ceases to grow further and 
stacking hybrids emerge on the exposed (111) face of the CD polymorph.

\subsection{Intermediate patches}

\begin{figure*}[h!]
    \centering
    \includegraphics[width=\linewidth]{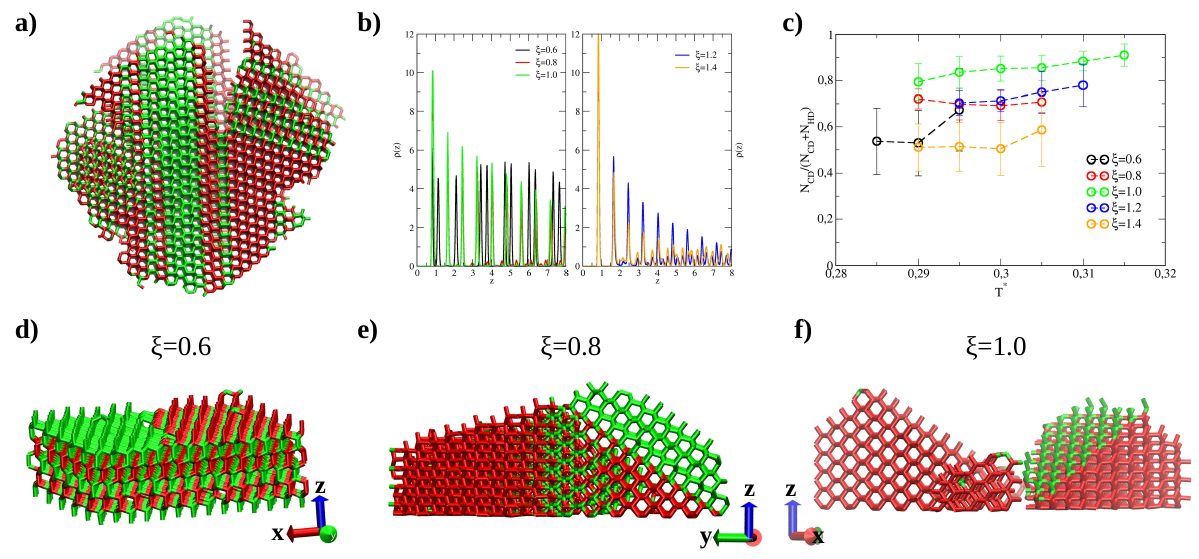}
    \caption{Part (a): Snapshots in the bulk system at $\rho=0.4$ and $T=0.285$. 
    Density profiles (b) and cubicity parameter (c) for the considered systems under different strengths of an applied external field. 
    Parts (d-f): Snapshots demonstrating different crystal growth mechanisms depending on the strength of the surface potential at $T=0.29$. 
    Red and green sticks correspond to the cubic and hexagonal diamond environments, respectively.}
    \label{fig:int}
\end{figure*}

The tetrahedral patchy particles with intermediate patches ($l=0.36\sigma$) 
in the bulk conditions assemble only into the CD/HD stacking hybrids in a one-component system \cite{Baran2023}.
As can be seen in Figure~\ref{fig:int}-a, the same ordered structures are formed in a two-component mixture. 
The ratio of diamond polymorphs also changes from replica to replica
as in the previous case ($l=0.34\sigma$). 

When the considered system is subjected to an external field below $\xi\le0.6$, the bulk-like networks are still observed. The characteristic bilayer structure in the density profiles can be seen in Figure~\ref{fig:int}-b, indicating the (0001) and (111) faces of HD and CD, respectively. 
Similarly, as in the case of narrow patches $(l=0.34\sigma)$, 
the external fields facilitate its growth, however, the grain boundaries 
are also visible (cf. Fig.~\ref{fig:int}-d). The cubicity parameter shown in Figure~\ref{fig:int}-c indicates that no polymorph is favoured over the other. 

\begin{figure}[h!]
    \centering
    \includegraphics[width=0.9\linewidth]{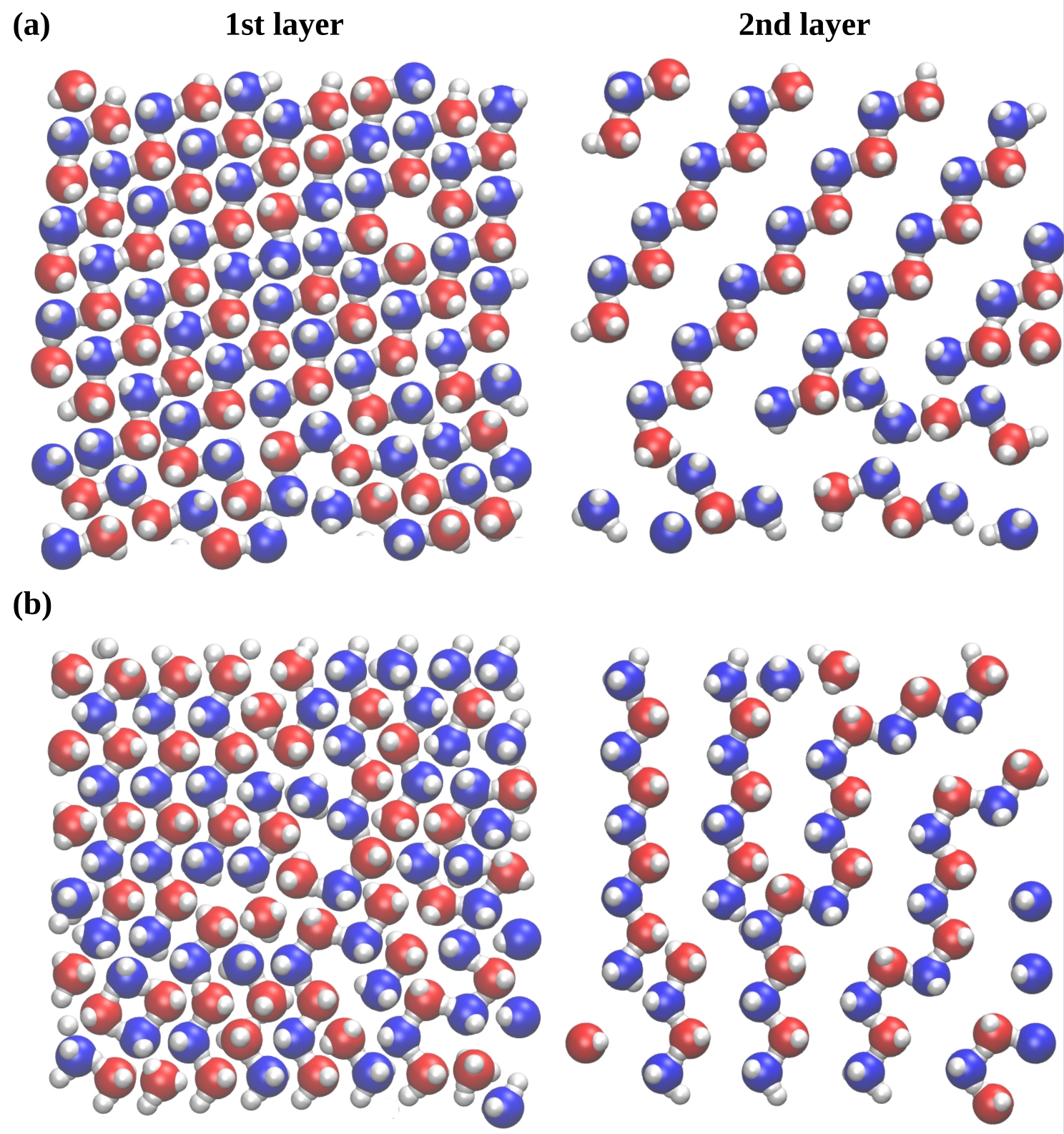}
    \caption{Structures of the first (left panel) and the second (right panel) adsorption layers for the cubic diamond which is formed 
    in the systems with potential $\xi=1.0$ (a) and $\xi=1.2$ (b)
    at $T=0.29$.
    Patchy particles of different species are marked in different colours. 
}
    \label{fig:int-layers}
\end{figure}

On the other hand, as the strength is increased to $\xi=0.8$, the bilayer vanishes and single peaks appear on the density profile. Therefore, the mechanism is essentially the same as that observed for the narrow patches ($l=0.34\sigma)$ where the patchy particles form a first adsorption layer with a triangular structure. This structure is commensurate with (110) face of the CD polymorph, thus leading to the favoured growth of this network as shown in Figure~\ref{fig:int}-e. 
However, the cubicity parameter varies in the range of $\approx0.7-0.8$ (cf. Fig.~\ref{fig:int}-c). This means that the selectivity has never been achieved with {\color{black}, however, the extent of particles identified as the CD polymorph is higher than those of the HD. Nevertheless, based on the argument about the frustrated antiferromagnetic triangular lattice, we anticipate that frustrations propagating in the lattice cannot allow for selectivity in the formation of a single polymorph. 
Even though the system initially nucleates into the CD polymorph, as the crystal layer becomes thicker (111) face of the CD polymorph becomes exposed. 
This is essentially the same mechanism as observed for the particles with narrow patches.} This can be readily seen in Figure~\ref{fig:int}-f but also in Figure~\ref{fig:int-layers}. In the former Figure, 
the formation of a monocrystal of CD has never been achieved and the
stacking hybrids emerge on the (111) face which is manifested by the presence of intermediate peaks on the density profile (cf. Fig.~\ref{fig:int}-b). 
Examining the adsorption layers shown in Figure~\ref{fig:int-layers} seems to confirm our suspicions. The first and second adsorption layers are almost identical for both systems with $\xi=1.0$ (Fig.~\ref{fig:int-layers}-a) and $\xi=1.2$ (Fig.~\ref{fig:int-layers}-b). 

{\color{black}
Therefore, we conclude that the simulation results obtained for the system
with the intermediate patch width ($l=0.36\sigma$) are almost identical
to those with narrow patches ($l=0.34\sigma$). The most prominent difference
is the change in the temperature scale towards higher values. 
In consequence, this can explain higher values of the cubicity parameter for $l=0.36\sigma$ as the thermal moves facilitate ordering
in developing crystals. Moreover, wider patches have less directional interactions, 
facilitating the alignment of particles into the crystalline network. 
}
\subsection{Wide patches}

\begin{figure*}[h!]
    \centering
    \includegraphics[width=0.85\linewidth]{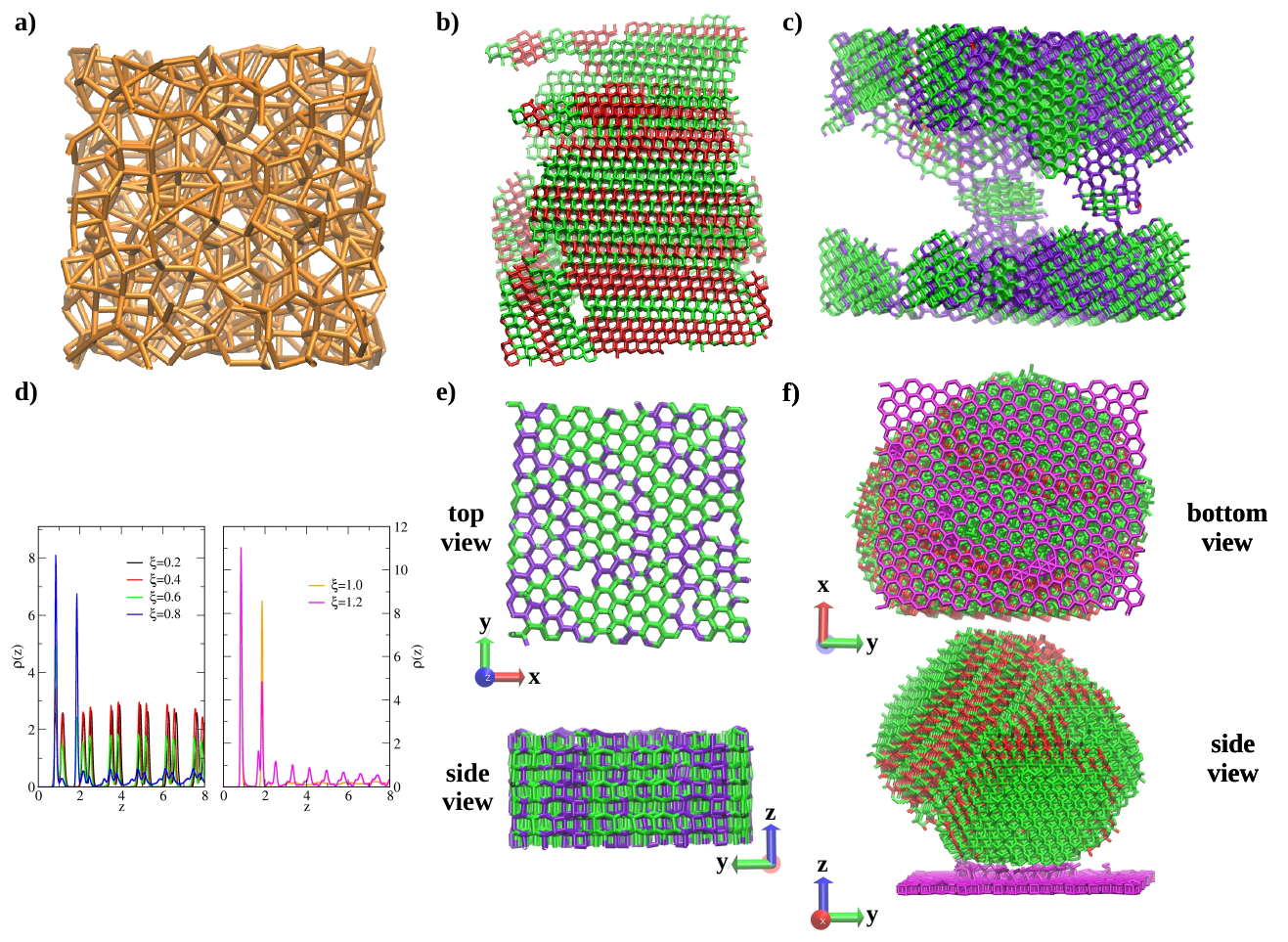}
    \caption{Snapshots in the bulk system for one-component system at $\rho=0.4$ and $T=0.37$ (a) and two-component systems 
    for two distinct replicas forming stacking hybrids (b) and 
     only hexagonal diamond (c) at $\rho=0.4$ and $T=0.37$. 
    Density profiles d) for the considered systems under different strengths of an applied external field. 
    Parts (e, f): Snapshots demonstrating different crystal growth mechanisms for the surface potential $\xi=0.4$ (e) and $\xi=0.8$ (f), both at $T=0.37$.
    Red, green, and purple sticks correspond to the cubic, hexagonal, and interfacial diamond environments, respectively.}
    \label{fig:wide}
\end{figure*}

One-component system comprising of tetrahedral particles with wide patches $(l=0.40\sigma$) forms disordered, glassy-like networks in the bulk (Fig.~\ref{fig:wide}-a). This is consistent with previous findings \cite{sanz, sanz1} reporting that the formation of diamond networks is hindered for tetrahedral patchy particles with patch's half-angle over $15^\circ$. On the other hand, in a 1:1 mixture, the formation of CD/HD stacking hybrids has been observed as can be seen in Figure~\ref{fig:wide}-b, similarly to Neophytou \textit{et al}. These authors found the proportion of the CD polymorph to be always higher than the HD one. Contrary to their results, in the present case, it has been found that some replicas form the HD crystals selectively (cf. Fig.~\ref{fig:wide}-c). However, notice that there is a significant number of molecules 
labelled as interfacial diamond (marked in purple, cf. Methods) connecting the molecules detected to be in a hexagonal diamond environment. 

For the systems subjected to external field potential up to a strength of 
$\xi=0.6$, a characteristic bilayer structure can be seen in the density profiles  (cf. Fig.~\ref{fig:wide}-d), similar to the systems with narrower patches. 
{\color{black} The arrangement of each kind of particles
and corresponding density profiles can be found in Figure~S3.
It has to be emphasized that the interactions between different species
are between the layers of a bilayer, similar to the systems with narrower patches.}

Surprisingly, structural analysis reveals the formation of only the HD polymorph for all the replicas (cf. Fig.~\ref{fig:wide}-e), 
instead of the stacking hybrids as has been the case for all the previously examined systems. 
{\color{black} We expect that the reason is the formation of many interfacial diamond molecules connecting the molecules identified as hexagonal diamonds. These particles tend to form characteristic four-membered rings which are flat. 
Such four- and six-membered rings are prerequisites for the formation of 
an s-III clathrate structure, however, its formation has never been observed. 
We anticipate that the competition between the formation of boat and chair conformations in a diamond crystal between the planar hexagon and also the appearance of flat four-membered rings are the reasons why the emergence of CD polymorph is hindered, resulting in the identification of particles belonging to only HD and interfacial diamond environments.}

\begin{figure}[h!]
    \centering
    \includegraphics[width=\linewidth]{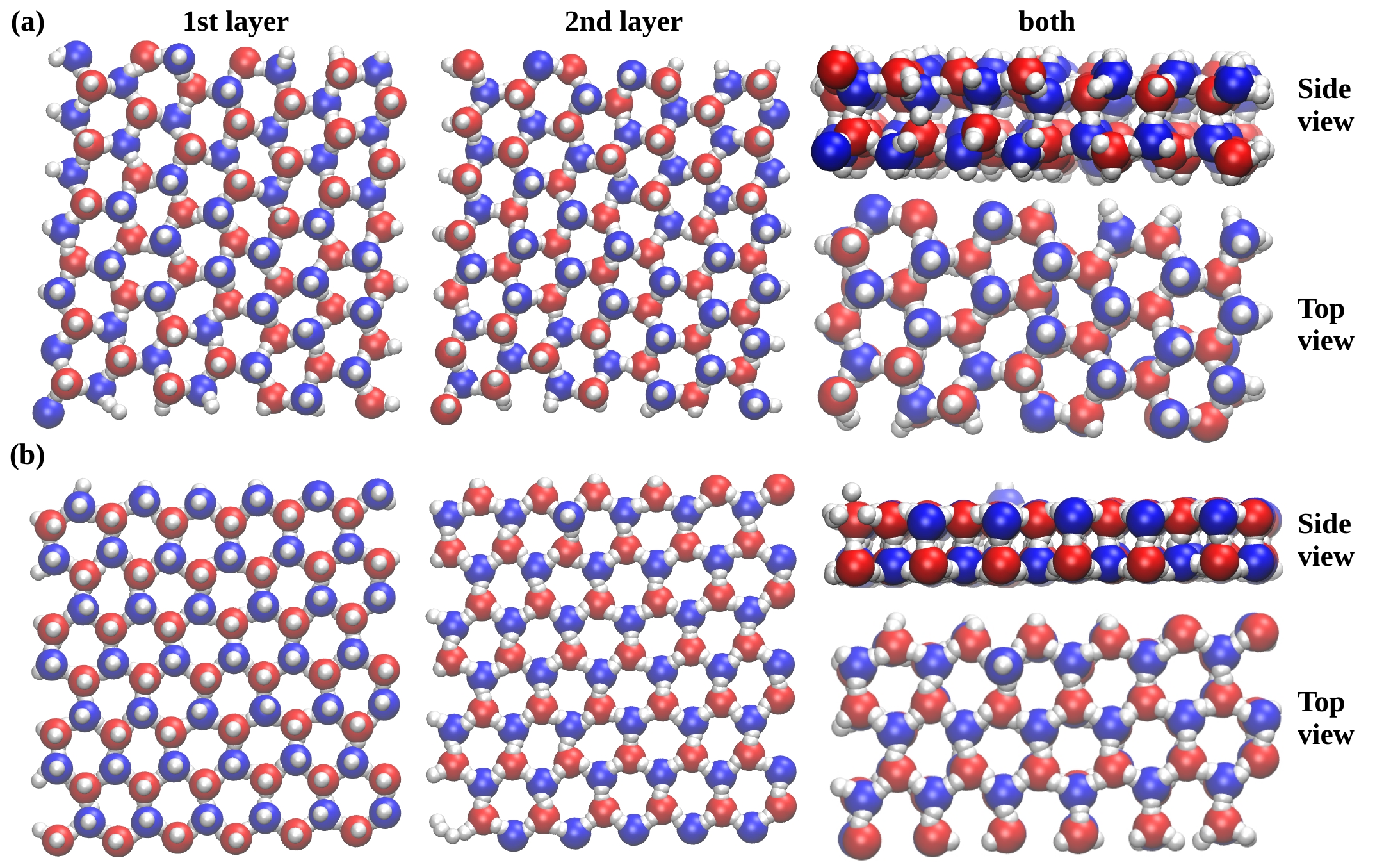}
    \caption{Structures of the first (left-hand side panel) and the second (middle panel) adsorption layers and how they are connected one with another (right-hand side panel) forming in the systems with surface potential $\xi=0.4$ (a) and $\xi=0.8$ (b). 
    Patchy particles of different species are marked in different colours. }
    \label{fig:wide-layers}
\end{figure}

As the external field is increased to $\xi=0.8$, the nucleation mechanism changes significantly. Instead of the bilayer structure, we observe only two single peaks and once the temperature drops to $T=0.37$, which is below the nucleation temperature for smaller values of $\xi$, 
some repeating peaks appear and their structure is diffused. 
We have found that in the first two adsorbed layers the patchy particles arrange
into a honeycomb structure. 
Recall that the bilayer observed for smaller values of $\xi$ resembles a hexagon 
as seen from the (0001) or (111) faces of HD and CD polymorphs, respectively.
Therefore, it seems that upon the increase in the external field, this bilayer structure flattens {\color{black} and particles are arranged into a honeycomb-like pattern, instead.} 
The comparison between a bilayer and a flat honeycomb structure can be found 
in Figure~\ref{fig:wide-layers}-a,b. 
In such a case, the tetrahedral patchy particles start behaving as 
a pseudo-trivalent since one active site of each particle 
is bound to form such a bilayer. 
On the other hand, the diffused peaks 
correspond to the formation of the bulk-like droplet growing on top of the bilayer
as can be seen in Figure~\ref{fig:wide}-f. 
{\color{black} 
Since the particles in the second and the consecutive layers are not interacting,
so the formed honeycomb bilayer is not wetted by the droplet growing on top of it. 
Therefore, the decrease in temperature is caused by the fact the 
system needs to reach a bulk nucleation temperature for this density. }

Recently, a case of such a colloidal graphene-like monolayer has been 
obtained experimentally \cite{Swinkels2023}, employing very similar strategy.
However, in the present case, tetrahedral particles interact only between different species. Such design results in the formation of two planar layers of the same structure
where there is no offset between the interconnected layers (compare right panels of Fig.~\ref{fig:wide-layers}-a,b), resembling boron nitride,
the so-called "inorganic graphite" \cite{azotek-boru1, azotek-boru2}.
Notice that, unlike the triangular lattice, such a two-dimensional 
honeycomb arrangement is not frustrated and essentially defect-free.

{\color{black}
As the external potential is increased to $\xi=1.0$ and $\xi=1.2$, we have observed that the density in the first adsorbing layers increases. 
It results in the honeycomb bilayer arrangement becoming disrupted
as shown in Figure~\ref{fig:wide-layers2}-a. It can be readily seen that some particles start filling the empty spaces and the planar honeycomb structure locally transforms into a triangular lattice, which resembles that formed for narrow ($l=0.34\sigma$) and intermediate ($l=0.36\sigma$) patches. The consequence of such behaviour is
that these two layers interact with the consecutive layers.
In fact, the crystalline droplet growing atop it originates 
from these "defects" which can be seen in Figure~\ref{fig:wide-layers2}-b.

\begin{figure}[h!]
    \centering
    \includegraphics[width=\linewidth]{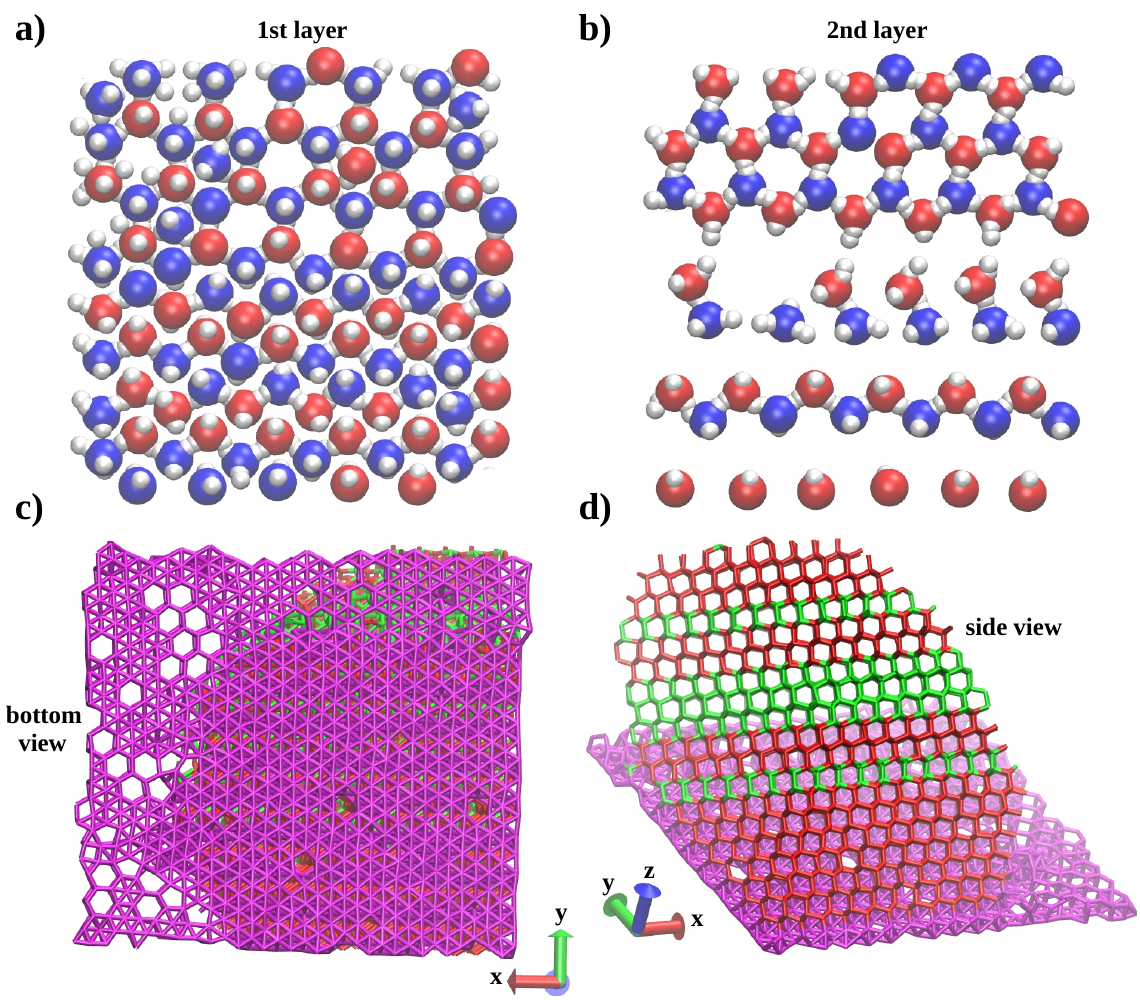}
    \caption{Structures of the first (a) and the second (b) adsorption layers that forms in the systems with a surface potential $\xi=0.8$ at $T=0.37$. 
    Patchy particles of different species are marked in different colours.
    Parts (c, d): snapshots demonstrating the crystal growth mechanism for these conditions. }
    \label{fig:wide-layers2}
\end{figure}

}

\section{Conclusions}

In this paper, we have demonstrated that the 1:1 mixture of 
tetrahedral patchy particles interacting only between distinct species
exhibits significant differences compared to the one-component system.
Such design resulted in the formation of only an even-membered ring 
was possible to promote the formation of diamond phases in a larger range
of the patch's width. Moreover, the formation of competing networks such
as s-II type clathrates or glasses was hindered.  

On the other hand, the selectivity of the formation of CD polymorph 
has never been achieved in a mixture as compared to the one-component systems.
Despite subjecting the system to a sufficiently strong external field which enforced the formation of a triangular lattice with a unit cell commensurate to the (110) face of CD polymorph, 
the frustrations in such an "antiferromagnetic-like" system lead to the emergence of several monocrystals and exposition of the (111) face of CD on which the stacking hybrids grow. 
{\color{black}
Despite that for intermediate patches ($l=0.36$) we observe
that the cubicity parameter reaches values as high as $\approx 0.7-0.8$ the selectivity cannot be achieved since the antiferromagnetic triangular lattice is frustrated.}
This conclusion is supported by smaller values of the cubicity parameter
for higher values of $\xi$ which has not been the case for one-component system \cite{Baran2023}.
{\color{black}
Another effect is related to the shift in the temperature scale 
to higher values as the patch's size (embedding distance $l$) is increased. 
Thermal moves facilitate the formation of well-developed crystals. In fact,
this is manifested by the comparison of the cubicity parameter
for the narrow ($l=0.34\sigma$) and intermediate ($l=0.36\sigma$) patches. 

On top of that, in systems with wide patches ($l=0.40\sigma$),
for a sufficiently strong surface potential $\xi$ the formation of a characteristic 
two planar graphene-like layers has been observed. Since there is no offset between them, such a setup resembles the boron nitride, the so-called "inorganic graphite" \cite{azotek-boru1, azotek-boru2} which is a promising candidate for many technological applications \cite{azotek-boru-app}. 

We expect that these findings can pave the way for the experimental realization
of diamond phases {\color{black}since the current simulation setup is not that different as compared to recent experimental reports \cite{Yamanaka2024, Swinkels2023}}. This study demonstrates why the two components are insufficient for the emergence of the CD polymorph selectively, even when the system is subjected to the external field potential. On the other hand, such a setup allows for the development
of (stacked) diamond phases in a broader range of conditions from tetrahedral patchy particles.

}

\section*{Author contributions}
 Dariusz Tarasewicz:
Formal analysis (equal); Writing - original draft (supporting). 
Edyta Raczy{\l\l}o: Formal analysis (equal).
Wojciech R{\.z}ysko: Formal analysis (equal); Writing - original draft (supporting). {\L}ukasz Baran: Formal analysis (equal); Investigation (lead); Writing - original draft (lead); Conceptualization (lead).

\section*{Conflicts of interest}
There are no conflicts to declare.
\section*{Acknowledgements}
This work was supported by the National Science Centre, Poland, under grant no. 2021/41/N/ST4/00437, PRELUDIUM 20. We gratefully acknowledge Polish high-performance computing infrastructure PLGrid (HPC Center: ACK Cyfronet AGH) for providing computer facilities and support within computational grant no. PLG/2024/017166.

\clearpage

\bibliography{biblio}

\providecommand{\latin}[1]{#1}
\makeatletter
\providecommand{\doi}
  {\begingroup\let\do\@makeother\dospecials
  \catcode`\{=1 \catcode`\}=2 \doi@aux}
\providecommand{\doi@aux}[1]{\endgroup\texttt{#1}}
\makeatother
\providecommand*\mcitethebibliography{\thebibliography}
\csname @ifundefined\endcsname{endmcitethebibliography}  {\let\endmcitethebibliography\endthebibliography}{}
\begin{mcitethebibliography}{46}
\providecommand*\natexlab[1]{#1}
\providecommand*\mciteSetBstSublistMode[1]{}
\providecommand*\mciteSetBstMaxWidthForm[2]{}
\providecommand*\mciteBstWouldAddEndPuncttrue
  {\def\EndOfBibitem{\unskip.}}
\providecommand*\mciteBstWouldAddEndPunctfalse
  {\let\EndOfBibitem\relax}
\providecommand*\mciteSetBstMidEndSepPunct[3]{}
\providecommand*\mciteSetBstSublistLabelBeginEnd[3]{}
\providecommand*\EndOfBibitem{}
\mciteSetBstSublistMode{f}
\mciteSetBstMaxWidthForm{subitem}{(\alph{mcitesubitemcount})}
\mciteSetBstSublistLabelBeginEnd
  {\mcitemaxwidthsubitemform\space}
  {\relax}
  {\relax}

\bibitem[Ravaine and Duguet(2017)Ravaine, and Duguet]{Duguet2017}
Ravaine,~S.; Duguet,~E. Synthesis and assembly of patchy particles: Recent progress and future prospects. \emph{Current Opinion in Colloid \& Interface Science} \textbf{2017}, \emph{30}, 45--53\relax
\mciteBstWouldAddEndPuncttrue
\mciteSetBstMidEndSepPunct{\mcitedefaultmidpunct}
{\mcitedefaultendpunct}{\mcitedefaultseppunct}\relax
\EndOfBibitem
\bibitem[Li \latin{et~al.}(2020)Li, Palis, Merindol, Majimel, Ravaine, and Duguet]{Duguet2020}
Li,~W.; Palis,~H.; Merindol,~R.; Majimel,~J.; Ravaine,~S.; Duguet,~E. Colloidal molecules and patchy particles: complementary concepts{,} synthesis and self-assembly. \emph{Chem. Soc. Rev.} \textbf{2020}, \emph{49}, 1955--1976\relax
\mciteBstWouldAddEndPuncttrue
\mciteSetBstMidEndSepPunct{\mcitedefaultmidpunct}
{\mcitedefaultendpunct}{\mcitedefaultseppunct}\relax
\EndOfBibitem
\bibitem[Poon(2004)]{Poon2004}
Poon,~W. Colloids as Big Atoms. \emph{Science} \textbf{2004}, \emph{304}, 830--831\relax
\mciteBstWouldAddEndPuncttrue
\mciteSetBstMidEndSepPunct{\mcitedefaultmidpunct}
{\mcitedefaultendpunct}{\mcitedefaultseppunct}\relax
\EndOfBibitem
\bibitem[Maldovan and Thomas(2004)Maldovan, and Thomas]{photonic_appl}
Maldovan,~M.; Thomas,~E.~L. Diamond-structured photonic crystals. \emph{Nature Materials} \textbf{2004}, \emph{3}, 593--600\relax
\mciteBstWouldAddEndPuncttrue
\mciteSetBstMidEndSepPunct{\mcitedefaultmidpunct}
{\mcitedefaultendpunct}{\mcitedefaultseppunct}\relax
\EndOfBibitem
\bibitem[Cersonsky \latin{et~al.}(2021)Cersonsky, Antonaglia, Dice, and Glotzer]{Cersonsky21}
Cersonsky,~R.~K.; Antonaglia,~J.; Dice,~B.~D.; Glotzer,~S.~C. The diversity of three-dimensional photonic crystals. \emph{Nature Communications} \textbf{2021}, \emph{12}, 2543\relax
\mciteBstWouldAddEndPuncttrue
\mciteSetBstMidEndSepPunct{\mcitedefaultmidpunct}
{\mcitedefaultendpunct}{\mcitedefaultseppunct}\relax
\EndOfBibitem
\bibitem[Bassani \latin{et~al.}(2024)Bassani, van Anders, Banin, Baranov, Chen, Dijkstra, Dimitriyev, Efrati, Faraudo, Gang, Gaston, Golestanian, Guerrero-Garcia, Gruenwald, Haji-Akbari, Ibáñez, Karg, Kraus, Lee, Van~Lehn, Macfarlane, Mognetti, Nikoubashman, Osat, Prezhdo, Rotskoff, Saiz, Shi, Skrabalak, Smalyukh, Tagliazucchi, Talapin, Tkachenko, Tretiak, Vaknin, Widmer-Cooper, Wong, Ye, Zhou, Rabani, Engel, and Travesset]{Travesset2024}
Bassani,~C.~L.; van Anders,~G.; Banin,~U.; Baranov,~D.; Chen,~Q.; Dijkstra,~M.; Dimitriyev,~M.~S.; Efrati,~E.; Faraudo,~J.; Gang,~O. \latin{et~al.}  Nanocrystal Assemblies: Current Advances and Open Problems. \emph{ACS Nano} \textbf{2024}, \emph{18}, 14791--14840, PMID: 38814908\relax
\mciteBstWouldAddEndPuncttrue
\mciteSetBstMidEndSepPunct{\mcitedefaultmidpunct}
{\mcitedefaultendpunct}{\mcitedefaultseppunct}\relax
\EndOfBibitem
\bibitem[Neophytou \latin{et~al.}(2021)Neophytou, Chakrabarti, and Sciortino]{Tetraedr_PNAS}
Neophytou,~A.; Chakrabarti,~D.; Sciortino,~F. Facile self-assembly of colloidal diamond from tetrahedral patchy particles via ring selection. \emph{Proceedings of the National Academy of Sciences} \textbf{2021}, \emph{118}, e2109776118\relax
\mciteBstWouldAddEndPuncttrue
\mciteSetBstMidEndSepPunct{\mcitedefaultmidpunct}
{\mcitedefaultendpunct}{\mcitedefaultseppunct}\relax
\EndOfBibitem
\bibitem[Noya \latin{et~al.}(2019)Noya, Zubieta, Pine, and Sciortino]{Noya2019}
Noya,~E.~G.; Zubieta,~I.; Pine,~D.~J.; Sciortino,~F. Assembly of clathrates from tetrahedral patchy colloids with narrow patches. \emph{The Journal of Chemical Physics} \textbf{2019}, \emph{151}, 094502\relax
\mciteBstWouldAddEndPuncttrue
\mciteSetBstMidEndSepPunct{\mcitedefaultmidpunct}
{\mcitedefaultendpunct}{\mcitedefaultseppunct}\relax
\EndOfBibitem
\bibitem[Neophytou \latin{et~al.}(2021)Neophytou, Manoharan, and Chakrabarti]{Nano_rods2021}
Neophytou,~A.; Manoharan,~V.~N.; Chakrabarti,~D. Self-Assembly of Patchy Colloidal Rods into Photonic Crystals Robust to Stacking Faults. \emph{ACS Nano} \textbf{2021}, \emph{15}, 2668--2678, PMID: 33448214\relax
\mciteBstWouldAddEndPuncttrue
\mciteSetBstMidEndSepPunct{\mcitedefaultmidpunct}
{\mcitedefaultendpunct}{\mcitedefaultseppunct}\relax
\EndOfBibitem
\bibitem[Noya \latin{et~al.}(2010)Noya, Vega, Doye, and Louis]{Noya1}
Noya,~E.~G.; Vega,~C.; Doye,~J. P.~K.; Louis,~A.~A. The stability of a crystal with diamond structure for patchy particles with tetrahedral symmetry. \emph{The Journal of Chemical Physics} \textbf{2010}, \emph{132}, 234511\relax
\mciteBstWouldAddEndPuncttrue
\mciteSetBstMidEndSepPunct{\mcitedefaultmidpunct}
{\mcitedefaultendpunct}{\mcitedefaultseppunct}\relax
\EndOfBibitem
\bibitem[Romano and Sciortino(2012)Romano, and Sciortino]{Romano2012}
Romano,~F.; Sciortino,~F. Patterning symmetry in the rational design of colloidal crystals. \emph{Nature Communications} \textbf{2012}, \emph{3}, 975\relax
\mciteBstWouldAddEndPuncttrue
\mciteSetBstMidEndSepPunct{\mcitedefaultmidpunct}
{\mcitedefaultendpunct}{\mcitedefaultseppunct}\relax
\EndOfBibitem
\bibitem[Romano \latin{et~al.}(2011)Romano, Sanz, and Sciortino]{sanz}
Romano,~F.; Sanz,~E.; Sciortino,~F. Crystallization of tetrahedral patchy particles in silico. \emph{The Journal of Chemical Physics} \textbf{2011}, \emph{134}, 174502\relax
\mciteBstWouldAddEndPuncttrue
\mciteSetBstMidEndSepPunct{\mcitedefaultmidpunct}
{\mcitedefaultendpunct}{\mcitedefaultseppunct}\relax
\EndOfBibitem
\bibitem[Romano \latin{et~al.}(2010)Romano, Sanz, and Sciortino]{sanz1}
Romano,~F.; Sanz,~E.; Sciortino,~F. Phase diagram of a tetrahedral patchy particle model for different interaction ranges. \emph{The Journal of Chemical Physics} \textbf{2010}, \emph{132}, 184501\relax
\mciteBstWouldAddEndPuncttrue
\mciteSetBstMidEndSepPunct{\mcitedefaultmidpunct}
{\mcitedefaultendpunct}{\mcitedefaultseppunct}\relax
\EndOfBibitem
\bibitem[Zhang \latin{et~al.}(2005)Zhang, Keys, Chen, and Glotzer]{zhang1}
Zhang; Keys,~A.~S.; Chen,~T.; Glotzer,~S.~C. Self-Assembly of Patchy Particles into Diamond Structures through Molecular Mimicry. \emph{Langmuir} \textbf{2005}, \emph{21}, 11547--11551, PMID: 16316077\relax
\mciteBstWouldAddEndPuncttrue
\mciteSetBstMidEndSepPunct{\mcitedefaultmidpunct}
{\mcitedefaultendpunct}{\mcitedefaultseppunct}\relax
\EndOfBibitem
\bibitem[Marin-Aguilar \latin{et~al.}(2022)Marin-Aguilar, Camerin, and Dijkstra]{dijkstra}
Marin-Aguilar,~S.; Camerin,~F.; Dijkstra,~M. {Guiding the self-assembly of colloidal diamond}. \emph{The Journal of Chemical Physics} \textbf{2022}, \emph{157}, 154503\relax
\mciteBstWouldAddEndPuncttrue
\mciteSetBstMidEndSepPunct{\mcitedefaultmidpunct}
{\mcitedefaultendpunct}{\mcitedefaultseppunct}\relax
\EndOfBibitem
\bibitem[Ducrot \latin{et~al.}(2017)Ducrot, He, Yi, and Pine]{Ducrot2017}
Ducrot,~{\'E}.; He,~M.; Yi,~G.-R.; Pine,~D.~J. Colloidal alloys with preassembled clusters and spheres. \emph{Nature Materials} \textbf{2017}, \emph{16}, 652--657\relax
\mciteBstWouldAddEndPuncttrue
\mciteSetBstMidEndSepPunct{\mcitedefaultmidpunct}
{\mcitedefaultendpunct}{\mcitedefaultseppunct}\relax
\EndOfBibitem
\bibitem[He \latin{et~al.}(2020)He, Gales, Ducrot, Gong, Yi, Sacanna, and Pine]{He2020}
He,~M.; Gales,~J.~P.; Ducrot,~{\'E}.; Gong,~Z.; Yi,~G.-R.; Sacanna,~S.; Pine,~D.~J. Colloidal diamond. \emph{Nature} \textbf{2020}, \emph{585}, 524--529\relax
\mciteBstWouldAddEndPuncttrue
\mciteSetBstMidEndSepPunct{\mcitedefaultmidpunct}
{\mcitedefaultendpunct}{\mcitedefaultseppunct}\relax
\EndOfBibitem
\bibitem[Liu \latin{et~al.}(2016)Liu, Tagawa, Xin, Wang, Emamy, Li, Yager, Starr, Tkachenko, and Gang]{Science2016}
Liu,~W.; Tagawa,~M.; Xin,~H.~L.; Wang,~T.; Emamy,~H.; Li,~H.; Yager,~K.~G.; Starr,~F.~W.; Tkachenko,~A.~V.; Gang,~O. Diamond family of nanoparticle superlattices. \emph{Science} \textbf{2016}, \emph{351}, 582--586\relax
\mciteBstWouldAddEndPuncttrue
\mciteSetBstMidEndSepPunct{\mcitedefaultmidpunct}
{\mcitedefaultendpunct}{\mcitedefaultseppunct}\relax
\EndOfBibitem
\bibitem[He \latin{et~al.}(2021)He, Gales, Shen, Kim, and Pine]{Pine2021}
He,~M.; Gales,~J.~P.; Shen,~X.; Kim,~M.~J.; Pine,~D.~J. Colloidal Particles with Triangular Patches. \emph{Langmuir} \textbf{2021}, \emph{37}, 7246--7253, PMID: 34081481\relax
\mciteBstWouldAddEndPuncttrue
\mciteSetBstMidEndSepPunct{\mcitedefaultmidpunct}
{\mcitedefaultendpunct}{\mcitedefaultseppunct}\relax
\EndOfBibitem
\bibitem[Posnjak \latin{et~al.}(2024)Posnjak, Yin, Butler, Bienek, Dass, Lee, Sharp, and Liedl]{Liedl2024}
Posnjak,~G.; Yin,~X.; Butler,~P.; Bienek,~O.; Dass,~M.; Lee,~S.; Sharp,~I.~D.; Liedl,~T. Diamond-lattice photonic crystals assembled from DNA origami. \emph{Science} \textbf{2024}, \emph{384}, 781--785\relax
\mciteBstWouldAddEndPuncttrue
\mciteSetBstMidEndSepPunct{\mcitedefaultmidpunct}
{\mcitedefaultendpunct}{\mcitedefaultseppunct}\relax
\EndOfBibitem
\bibitem[Liu \latin{et~al.}(2024)Liu, Matthies, Russo, Rovigatti, Narayanan, Diep, McKeen, Gang, Stephanopoulos, Sciortino, Yan, Romano, and Šulc]{Rovigatti2024}
Liu,~H.; Matthies,~M.; Russo,~J.; Rovigatti,~L.; Narayanan,~R.~P.; Diep,~T.; McKeen,~D.; Gang,~O.; Stephanopoulos,~N.; Sciortino,~F. \latin{et~al.}  Inverse design of a pyrochlore lattice of DNA origami through model-driven experiments. \emph{Science} \textbf{2024}, \emph{384}, 776--781\relax
\mciteBstWouldAddEndPuncttrue
\mciteSetBstMidEndSepPunct{\mcitedefaultmidpunct}
{\mcitedefaultendpunct}{\mcitedefaultseppunct}\relax
\EndOfBibitem
\bibitem[Falenty \latin{et~al.}(2014)Falenty, Hansen, and Kuhs]{Falenty2014}
Falenty,~A.; Hansen,~T.~C.; Kuhs,~W.~F. Formation and properties of ice XVI obtained by emptying a type sII clathrate hydrate. \emph{Nature} \textbf{2014}, \emph{516}, 231--233\relax
\mciteBstWouldAddEndPuncttrue
\mciteSetBstMidEndSepPunct{\mcitedefaultmidpunct}
{\mcitedefaultendpunct}{\mcitedefaultseppunct}\relax
\EndOfBibitem
\bibitem[Lin \latin{et~al.}(2017)Lin, Lee, Sun, Spellings, Engel, Glotzer, and Mirkin]{Glotzer_klatrat}
Lin,~H.; Lee,~S.; Sun,~L.; Spellings,~M.; Engel,~M.; Glotzer,~S.~C.; Mirkin,~C.~A. Clathrate colloidal crystals. \emph{Science} \textbf{2017}, \emph{355}, 931--935\relax
\mciteBstWouldAddEndPuncttrue
\mciteSetBstMidEndSepPunct{\mcitedefaultmidpunct}
{\mcitedefaultendpunct}{\mcitedefaultseppunct}\relax
\EndOfBibitem
\bibitem[Lee \latin{et~al.}(2023)Lee, Vo, and Glotzer]{Glotzer_klatrat2}
Lee,~S.; Vo,~T.; Glotzer,~S.~C. Entropy compartmentalization stabilizes open host--guest colloidal clathrates. \emph{Nature Chemistry} \textbf{2023}, \emph{15}, 905--912\relax
\mciteBstWouldAddEndPuncttrue
\mciteSetBstMidEndSepPunct{\mcitedefaultmidpunct}
{\mcitedefaultendpunct}{\mcitedefaultseppunct}\relax
\EndOfBibitem
\bibitem[Davies \latin{et~al.}(2021)Davies, Fitzner, and Michaelides]{IceIc_Michaelides}
Davies,~M.~B.; Fitzner,~M.; Michaelides,~A. Routes to cubic ice through heterogeneous nucleation. \emph{Proceedings of the National Academy of Sciences} \textbf{2021}, \emph{118}, e2025245118\relax
\mciteBstWouldAddEndPuncttrue
\mciteSetBstMidEndSepPunct{\mcitedefaultmidpunct}
{\mcitedefaultendpunct}{\mcitedefaultseppunct}\relax
\EndOfBibitem
\bibitem[Nandi \latin{et~al.}(2018)Nandi, Burnham, and English]{Niall2018}
Nandi,~P.~K.; Burnham,~C.~J.; English,~N.~J. Electro-nucleation of water nano-droplets in No Man{'}s Land to fault-free ice Ic. \emph{Phys. Chem. Chem. Phys.} \textbf{2018}, \emph{20}, 8042--8053\relax
\mciteBstWouldAddEndPuncttrue
\mciteSetBstMidEndSepPunct{\mcitedefaultmidpunct}
{\mcitedefaultendpunct}{\mcitedefaultseppunct}\relax
\EndOfBibitem
\bibitem[Koga \latin{et~al.}(2001)Koga, Gao, Tanaka, and Zeng]{Koga2001}
Koga,~K.; Gao,~G.~T.; Tanaka,~H.; Zeng,~X.~C. Formation of ordered ice nanotubes inside carbon nanotubes. \emph{Nature} \textbf{2001}, \emph{412}, 802--805\relax
\mciteBstWouldAddEndPuncttrue
\mciteSetBstMidEndSepPunct{\mcitedefaultmidpunct}
{\mcitedefaultendpunct}{\mcitedefaultseppunct}\relax
\EndOfBibitem
\bibitem[Koga \latin{et~al.}(2002)Koga, Gao, Tanaka, and Zeng]{Koga2002}
Koga,~K.; Gao,~G.; Tanaka,~H.; Zeng,~X. How does water freeze inside carbon nanotubes? \emph{Physica A: Statistical Mechanics and its Applications} \textbf{2002}, \emph{314}, 462--469, Horizons in Complex Systems\relax
\mciteBstWouldAddEndPuncttrue
\mciteSetBstMidEndSepPunct{\mcitedefaultmidpunct}
{\mcitedefaultendpunct}{\mcitedefaultseppunct}\relax
\EndOfBibitem
\bibitem[Araujo \latin{et~al.}(2023)Araujo, Janssen, Barois, Boffetta, Cohen, Corbetta, Dauchot, Dijkstra, Durham, Dussutour, Garnier, Gelderblom, Golestanian, Isa, Koenderink, Löwen, Metzler, Polin, Royall, Šarić, Sengupta, Sykes, Trianni, Tuval, Vogel, Yeomans, Zuriguel, Marin, and Volpe]{dijkstra_perspective}
Araujo,~N. A.~M.; Janssen,~L. M.~C.; Barois,~T.; Boffetta,~G.; Cohen,~I.; Corbetta,~A.; Dauchot,~O.; Dijkstra,~M.; Durham,~W.~M.; Dussutour,~A. \latin{et~al.}  Steering self-organisation through confinement. \emph{Soft Matter} \textbf{2023}, \emph{19}, 1695--1704\relax
\mciteBstWouldAddEndPuncttrue
\mciteSetBstMidEndSepPunct{\mcitedefaultmidpunct}
{\mcitedefaultendpunct}{\mcitedefaultseppunct}\relax
\EndOfBibitem
\bibitem[Baran \latin{et~al.}(2023)Baran, Tarasewicz, Kami{\'n}ski, and R{\.z}ysko]{Baran2023}
Baran,~{\L}.; Tarasewicz,~D.; Kami{\'n}ski,~D.~M.; R{\.z}ysko,~W. Pursuing colloidal diamonds. \emph{Nanoscale} \textbf{2023}, \emph{15}, 10623--10633\relax
\mciteBstWouldAddEndPuncttrue
\mciteSetBstMidEndSepPunct{\mcitedefaultmidpunct}
{\mcitedefaultendpunct}{\mcitedefaultseppunct}\relax
\EndOfBibitem
\bibitem[Baran \latin{et~al.}(2024)Baran, Tarasewicz, and R{\.z}ysko]{Baran2024}
Baran,~{\L}.; Tarasewicz,~D.; R{\.z}ysko,~W. Interplay between the Formation of Colloidal Clathrate and Cubic Diamond Crystals. \emph{The Journal of Physical Chemistry B} \textbf{2024}, \emph{128}, 5792--5801, PMID: 38832806\relax
\mciteBstWouldAddEndPuncttrue
\mciteSetBstMidEndSepPunct{\mcitedefaultmidpunct}
{\mcitedefaultendpunct}{\mcitedefaultseppunct}\relax
\EndOfBibitem
\bibitem[Wang \latin{et~al.}(2012)Wang, Wang, Breed, Manoharan, Feng, Hollingsworth, Weck, and Pine]{Wang2012}
Wang,~Y.; Wang,~Y.; Breed,~D.~R.; Manoharan,~V.~N.; Feng,~L.; Hollingsworth,~A.~D.; Weck,~M.; Pine,~D.~J. Colloids with valence and specific directional bonding. \emph{Nature} \textbf{2012}, \emph{491}, 51--55\relax
\mciteBstWouldAddEndPuncttrue
\mciteSetBstMidEndSepPunct{\mcitedefaultmidpunct}
{\mcitedefaultendpunct}{\mcitedefaultseppunct}\relax
\EndOfBibitem
\bibitem[Rogers \latin{et~al.}(2016)Rogers, Shih, and Manoharan]{Rogers2016}
Rogers,~W.~B.; Shih,~W.~M.; Manoharan,~V.~N. Using DNA to program the self-assembly of colloidal nanoparticles and microparticles. \emph{Nature Reviews Materials} \textbf{2016}, \emph{1}, 16008\relax
\mciteBstWouldAddEndPuncttrue
\mciteSetBstMidEndSepPunct{\mcitedefaultmidpunct}
{\mcitedefaultendpunct}{\mcitedefaultseppunct}\relax
\EndOfBibitem
\bibitem[Vasilyev \latin{et~al.}(2015)Vasilyev, Klumov, and Tkachenko]{chromatic2015}
Vasilyev,~O.~A.; Klumov,~B.~A.; Tkachenko,~A.~V. Chromatic patchy particles: Effects of specific interactions on liquid structure. \emph{Phys. Rev. E} \textbf{2015}, \emph{92}, 012308\relax
\mciteBstWouldAddEndPuncttrue
\mciteSetBstMidEndSepPunct{\mcitedefaultmidpunct}
{\mcitedefaultendpunct}{\mcitedefaultseppunct}\relax
\EndOfBibitem
\bibitem[Toxvaerd and Dyre(2011)Toxvaerd, and Dyre]{toxvaerd}
Toxvaerd,~S.; Dyre,~J.~C. {Communication: Shifted forces in molecular dynamics}. \emph{The Journal of Chemical Physics} \textbf{2011}, \emph{134}, 081102\relax
\mciteBstWouldAddEndPuncttrue
\mciteSetBstMidEndSepPunct{\mcitedefaultmidpunct}
{\mcitedefaultendpunct}{\mcitedefaultseppunct}\relax
\EndOfBibitem
\bibitem[Nguyen and Molinero(2015)Nguyen, and Molinero]{chill+}
Nguyen,~A.~H.; Molinero,~V. Identification of Clathrate Hydrates, Hexagonal Ice, Cubic Ice, and Liquid Water in Simulations: the CHILL+ Algorithm. \emph{The Journal of Physical Chemistry B} \textbf{2015}, \emph{119}, 9369--9376, PMID: 25389702\relax
\mciteBstWouldAddEndPuncttrue
\mciteSetBstMidEndSepPunct{\mcitedefaultmidpunct}
{\mcitedefaultendpunct}{\mcitedefaultseppunct}\relax
\EndOfBibitem
\bibitem[Thompson \latin{et~al.}(2022)Thompson, Aktulga, Berger, Bolintineanu, Brown, Crozier, in~'t Veld, Kohlmeyer, Moore, Nguyen, Shan, Stevens, Tranchida, Trott, and Plimpton]{LAMMPS}
Thompson,~A.~P.; Aktulga,~H.~M.; Berger,~R.; Bolintineanu,~D.~S.; Brown,~W.~M.; Crozier,~P.~S.; in~'t Veld,~P.~J.; Kohlmeyer,~A.; Moore,~S.~G.; Nguyen,~T.~D. \latin{et~al.}  {LAMMPS} - a flexible simulation tool for particle-based materials modeling at the atomic, meso, and continuum scales. \emph{Comp. Phys. Comm.} \textbf{2022}, \emph{271}, 108171\relax
\mciteBstWouldAddEndPuncttrue
\mciteSetBstMidEndSepPunct{\mcitedefaultmidpunct}
{\mcitedefaultendpunct}{\mcitedefaultseppunct}\relax
\EndOfBibitem
\bibitem[MacDougall \latin{et~al.}(2011)MacDougall, Gout, Zarestky, Ehlers, Podlesnyak, McGuire, Mandrus, and Nagler]{antiferromagnet1}
MacDougall,~G.~J.; Gout,~D.; Zarestky,~J.~L.; Ehlers,~G.; Podlesnyak,~A.; McGuire,~M.~A.; Mandrus,~D.; Nagler,~S.~E. Kinetically inhibited order in a diamond-lattice antiferromagnet. \emph{Proceedings of the National Academy of Sciences} \textbf{2011}, \emph{108}, 15693--15698\relax
\mciteBstWouldAddEndPuncttrue
\mciteSetBstMidEndSepPunct{\mcitedefaultmidpunct}
{\mcitedefaultendpunct}{\mcitedefaultseppunct}\relax
\EndOfBibitem
\bibitem[Oitmaa(2019)]{antiferromagnet2}
Oitmaa,~J. Frustrated diamond lattice antiferromagnet. \emph{Phys. Rev. B} \textbf{2019}, \emph{99}, 134407\relax
\mciteBstWouldAddEndPuncttrue
\mciteSetBstMidEndSepPunct{\mcitedefaultmidpunct}
{\mcitedefaultendpunct}{\mcitedefaultseppunct}\relax
\EndOfBibitem
\bibitem[Moessner and Ramirez(2006)Moessner, and Ramirez]{frustration}
Moessner,~R.; Ramirez,~A.~P. {Geometrical frustration}. \emph{Physics Today} \textbf{2006}, \emph{59}, 24--29\relax
\mciteBstWouldAddEndPuncttrue
\mciteSetBstMidEndSepPunct{\mcitedefaultmidpunct}
{\mcitedefaultendpunct}{\mcitedefaultseppunct}\relax
\EndOfBibitem
\bibitem[Swinkels \latin{et~al.}(2023)Swinkels, Gong, Sacanna, Noya, and Schall]{Swinkels2023}
Swinkels,~P. J.~M.; Gong,~Z.; Sacanna,~S.; Noya,~E.~G.; Schall,~P. Visualizing defect dynamics by assembling the colloidal graphene lattice. \emph{Nature Communications} \textbf{2023}, \emph{14}, 1524\relax
\mciteBstWouldAddEndPuncttrue
\mciteSetBstMidEndSepPunct{\mcitedefaultmidpunct}
{\mcitedefaultendpunct}{\mcitedefaultseppunct}\relax
\EndOfBibitem
\bibitem[Kawaguchi \latin{et~al.}(2008)Kawaguchi, Kuroda, and Muramatsu]{azotek-boru1}
Kawaguchi,~M.; Kuroda,~S.; Muramatsu,~Y. Electronic structure and intercalation chemistry of graphite-like layered material with a composition of BC6N. \emph{Journal of Physics and Chemistry of Solids} \textbf{2008}, \emph{69}, 1171--1178, 14th International Symposium on Intercalation Compounds\relax
\mciteBstWouldAddEndPuncttrue
\mciteSetBstMidEndSepPunct{\mcitedefaultmidpunct}
{\mcitedefaultendpunct}{\mcitedefaultseppunct}\relax
\EndOfBibitem
\bibitem[Ba \latin{et~al.}(2017)Ba, Jiang, Cheng, Bao, Xuan, Sun, Liu, Xie, Wu, and Sun]{azotek-boru2}
Ba,~K.; Jiang,~W.; Cheng,~J.; Bao,~J.; Xuan,~N.; Sun,~Y.; Liu,~B.; Xie,~A.; Wu,~S.; Sun,~Z. Chemical and Bandgap Engineering in Monolayer Hexagonal Boron Nitride. \emph{Scientific Reports} \textbf{2017}, \emph{7}, 45584\relax
\mciteBstWouldAddEndPuncttrue
\mciteSetBstMidEndSepPunct{\mcitedefaultmidpunct}
{\mcitedefaultendpunct}{\mcitedefaultseppunct}\relax
\EndOfBibitem
\bibitem[Vatanpour \latin{et~al.}(2021)Vatanpour, Naziri~Mehrabani, Keskin, Arabi, Zeytuncu, and Koyuncu]{azotek-boru-app}
Vatanpour,~V.; Naziri~Mehrabani,~S.~A.; Keskin,~B.; Arabi,~N.; Zeytuncu,~B.; Koyuncu,~I. A Comprehensive Review on the Applications of Boron Nitride Nanomaterials in Membrane Fabrication and Modification. \emph{Industrial \& Engineering Chemistry Research} \textbf{2021}, \emph{60}, 13391--13424\relax
\mciteBstWouldAddEndPuncttrue
\mciteSetBstMidEndSepPunct{\mcitedefaultmidpunct}
{\mcitedefaultendpunct}{\mcitedefaultseppunct}\relax
\EndOfBibitem
\bibitem[Fujita \latin{et~al.}(2024)Fujita, Toyotama, Okuzono, Niinomi, and Yamanaka]{Yamanaka2024}
Fujita,~M.; Toyotama,~A.; Okuzono,~T.; Niinomi,~H.; Yamanaka,~J. Formation of two-dimensional diamond-like colloidal crystals using layer-by-layer electrostatic self-assembly. \emph{Soft Matter} \textbf{2024}, \emph{20}, 985--992\relax
\mciteBstWouldAddEndPuncttrue
\mciteSetBstMidEndSepPunct{\mcitedefaultmidpunct}
{\mcitedefaultendpunct}{\mcitedefaultseppunct}\relax
\EndOfBibitem
\end{mcitethebibliography}
\bibstyle{achemso}

\clearpage


\section*{Supplementary material}

# Supplementary information.

Dariusz Tarasewicz, Edyta Raczyłło, Wojciech Rżysko, and Łukasz Baran

Arrangement of the "narrow" patchy particles ($l = 0.34\sigma$) of each kind in the first bilayer and corresponding density profiles are shown in Figures 1 2 for two distinct replicas. For the first case (Figure 1), one can see two domains of particles A and B in a single layer of the bilayer. Those domains have triangular symmetry, and the distance between particles of the same kind is $r \approx 1.56$, indicating that the domains are not densely packed. For the second replica (Figure 2), the same symmetry was observed in the each of the layers of the bilayer. However, the same structure is formed by a single component.

In the systems with the "wide" patches ($l = 0.40\sigma$) that are subjected to a weak external field, we observe that the bilayer's structure is essentially the same as in the case of the narrow patches (Figure 3). The only difference is the increased distance between the particles of the same kind which is $r \approx 1.65$. Figure

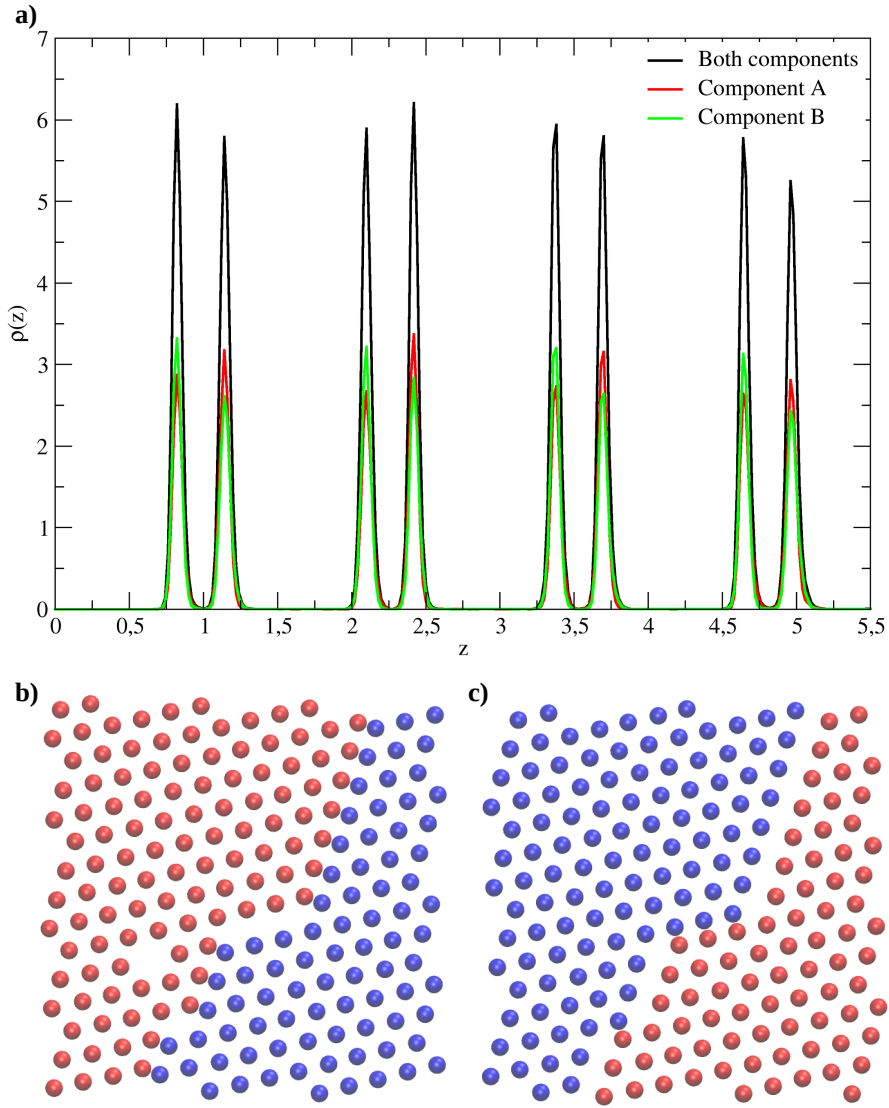

Figure 1: Part (a): total density profiles and for each of the components. Parts (b-c): snapshots of each of the layers forming the bilayer at distances $0.5 < z < 1.0$ and $1.0 < z < 1.5$, respectively. Patchy particles of different species are in different colors.

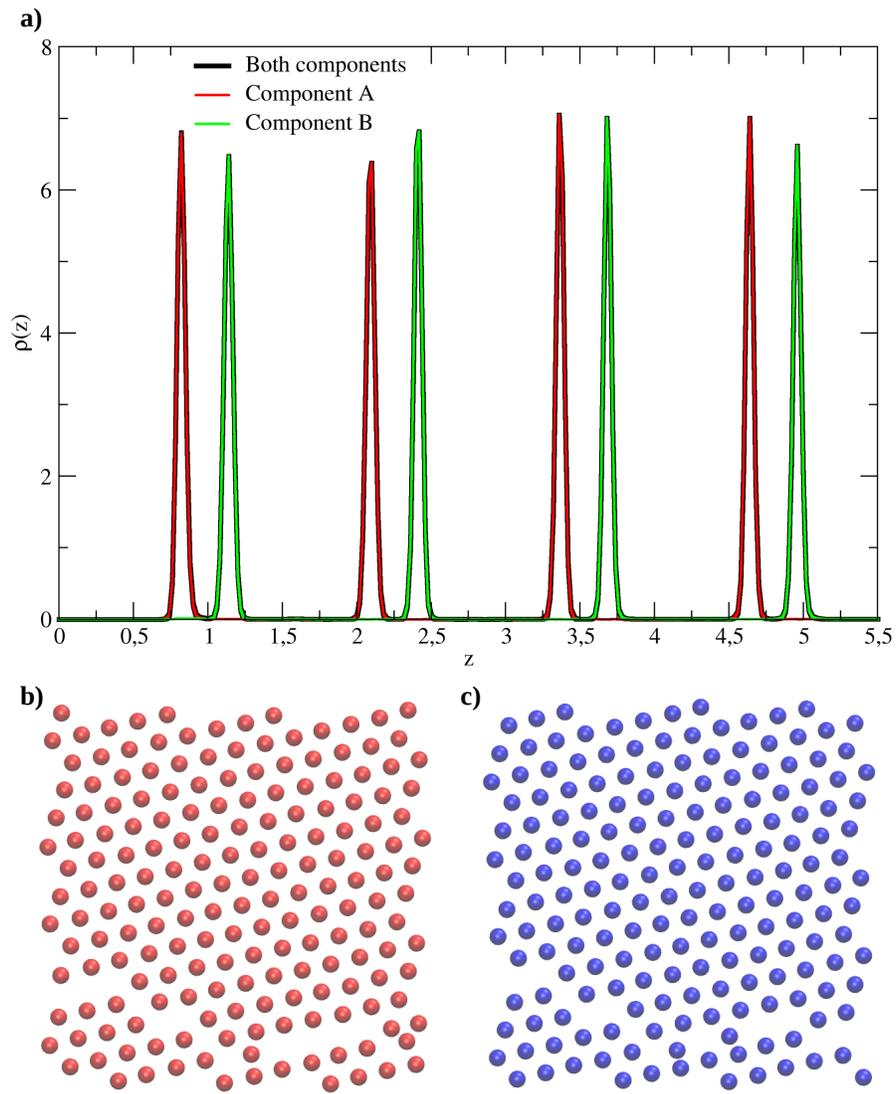

Figure 2: Part (a): total density profiles and for each of the components. Parts (b-c): snapshots of each of the layers forming the bilayer at distances $0.5 < z < 1.0$ and $1.0 < z < 1.5$, respectively. Patchy particles of different species are in different colors.

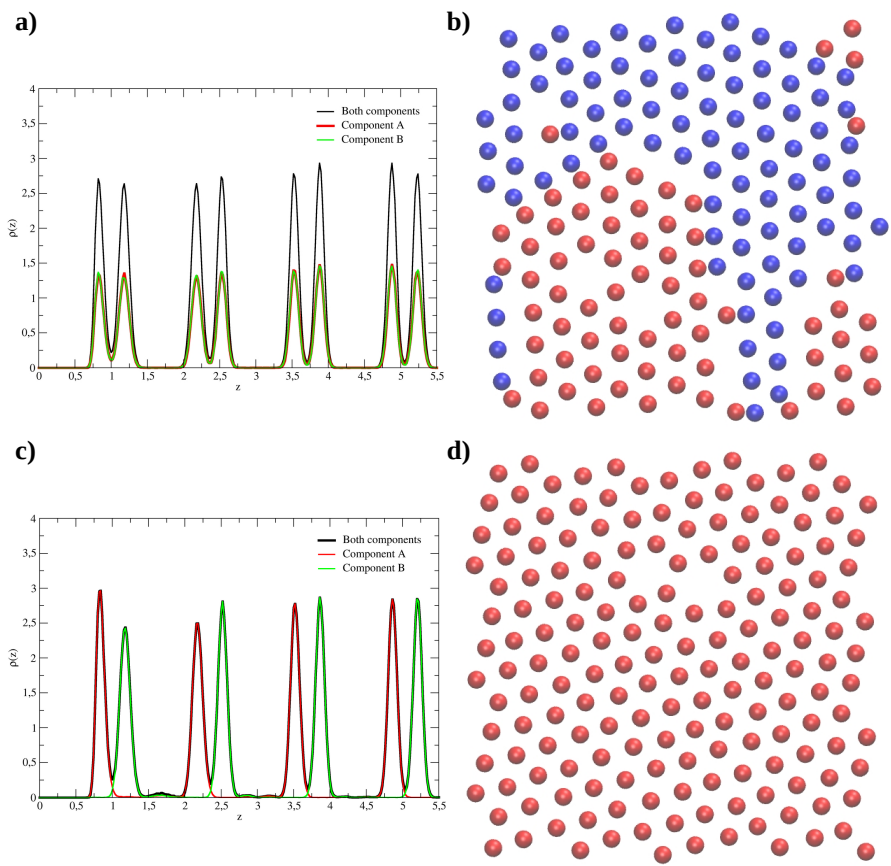

Figure 3: Parts (a) and (c): total density profiles and for each of the components. Parts (b) and (d): snapshots of a single layer of the bilayer at distances $0.5 < z < 1.0$ for two distinct replicas. Patchy particles of different species are in different colors.



\end{document}